\documentclass[sigconf]{acmart}
\acmConference[Draft]{arxiv}{December 2023}{}
\setcopyright{acmcopyright}
\copyrightyear{2023}
\acmYear{2023}
\acmDOI{XXXXXXX.XXXXXXX}
\acmISBN{978-1-4503-XXXX-X/18/06}

\acmSubmissionID{123-A56-BU3}

\usepackage{booktabs}
\usepackage{soul}
\usepackage{tabularx}
\usepackage{tabulary}
\usepackage[para,online,flushleft]{threeparttable}
\usepackage{tikz}
\usetikzlibrary{shapes.misc, positioning}
\usetikzlibrary{tikzmark}
\usepackage{float}

\usepackage{xcolor}

\usepackage{tablefootnote} 
\usepackage{enumitem}
\usepackage{subcaption}
\usepackage{multirow}
\usepackage[capitalize]{cleveref}

\usepackage{pifont}

\newcommand{\participants}{26}

\newcommand{\bob}{P1}

\newcommand{\alice}{P2}
\newcommand{\sally}{P3}
\newcommand{\steve}{P4}
\newcommand{\samim}{P5}

\newcommand{\ironman}{P7}
\newcommand{\batman}{P8}
\newcommand{\reddragon}{P9}
\newcommand{\thomas}{P10}
\newcommand{\thepope}{P11}
\newcommand{\ericbob}{P12}
\newcommand{\akond}{P14}
\newcommand{\nischal}{P15}
\newcommand{\brown}{P16}
\newcommand{\keagan}{P18}
\newcommand{\mahnaz}{P20}
\newcommand{\austin}{P22}
\newcommand{\barik}{P23}
\newcommand{\mickey}{P24}
\newcommand{\stanley}{P25}

\definecolor{jccolor}{rgb}{0.1,0.7,0.8}
\definecolor{vlcolor}{rgb}{0.9,0.1,0.1}
\definecolor{cpcolor}{rgb}{0.3,0.3,0.7}
\definecolor{ascolor}{RGB}{163,96,50}
\definecolor{ahcolor}{rgb}{0.36, 0.54, 0.66}

\definecolor{lightgray}{gray}{0.9}
\sethlcolor{lightgray}

\newcommand{\CalloutFigure}[1]{Fig.~\ref{#1}} 

\begin{document}

\title{Building Your Own Product Copilot: Challenges, \\Opportunities, and Needs}

\author{Chris Parnin, Gustavo Soares, Rahul Pandita, Sumit Gulwani, Jessica Rich, Austin Z. Henley}
\email{{chrisparnin, gustavo.soares}@microsoft.com, rahulpandita@github.com, {sumitg, jessrich, austinhenley}@microsoft.com}
\affiliation{%
  \institution{Microsoft, GitHub Inc.}
  \country{USA}
}

\begin{abstract}
A race is underway to embed advanced AI capabilities into products. These product ``copilots'' enable users to ask questions in natural language and receive relevant responses that are specific to the user's context. In fact, virtually every large technology company is looking to add these capabilities to their software products. However, for most software engineers, this is often their first encounter with integrating AI-powered technology. Furthermore, software engineering processes and tools have not caught up with the challenges and scale involved with building AI-powered applications. In this work, we present the findings of an interview study with 26 professional software engineers responsible for building product copilots at various companies. From our interviews, we found pain points at every step of the engineering process and the challenges that strained existing development practices. We then conducted group brainstorming sessions to collaborative on opportunities and tool designs for the broader software engineering community.
\end{abstract}




\maketitle

\section{Introduction}

So, you want to build a copilot? You are not alone. In the past year, a race has been underway to embed advanced AI capabilities into products and sometimes entire portfolios. Often, these come in the form of a conversational agent powered by large-language models (LLMs) and assist a user as a \emph{copilot} in performing their tasks. For example, Salesforce recently announced Einstein Copilot, which will ``assist users within their flow of work, enabling them to ask questions in natural language and receive relevant and trustworthy answers that are grounded in secure proprietary company data from Data Cloud.\footnote{\url{https://www.salesforce.com/news/press-releases/2023/09/12/salesforce-platform-news-dreamforce/}}.'' Virtually every large technology company is looking to add similar capabilities to their software products.

However, for most software engineers, this is often their first encounter with using and integrating AI-powered technology. Furthermore, software engineering processes and tools have not caught up with the challenges and scale involved with building AI-powered applications. Many questions, such as what is the best way to design and manage prompts, gather context about the application, manage conversational state, allow and provide agency to the AI copilot, test and verify end-to-end workflows, and organize teams to put it all together.

In this paper, we present the findings of an interview study with \participants{} professional software engineers responsible for building product copilots at various companies. From our interviews, we found pain points at every step of the engineering process and the challenges that strained existing development practices. In particular, prompt engineering and testing are extremely time-consuming and resource-constrained. Software engineers desire comprehensive tooling and best practices, which are yet to be defined. We then conducted group brainstorming sessions to collaboratively review possible opportunities and tool designs to address these challenges.

\section{Background}

\begin{figure*}
\subfloat[GitHub Copilot]{\includegraphics[width = 2in]{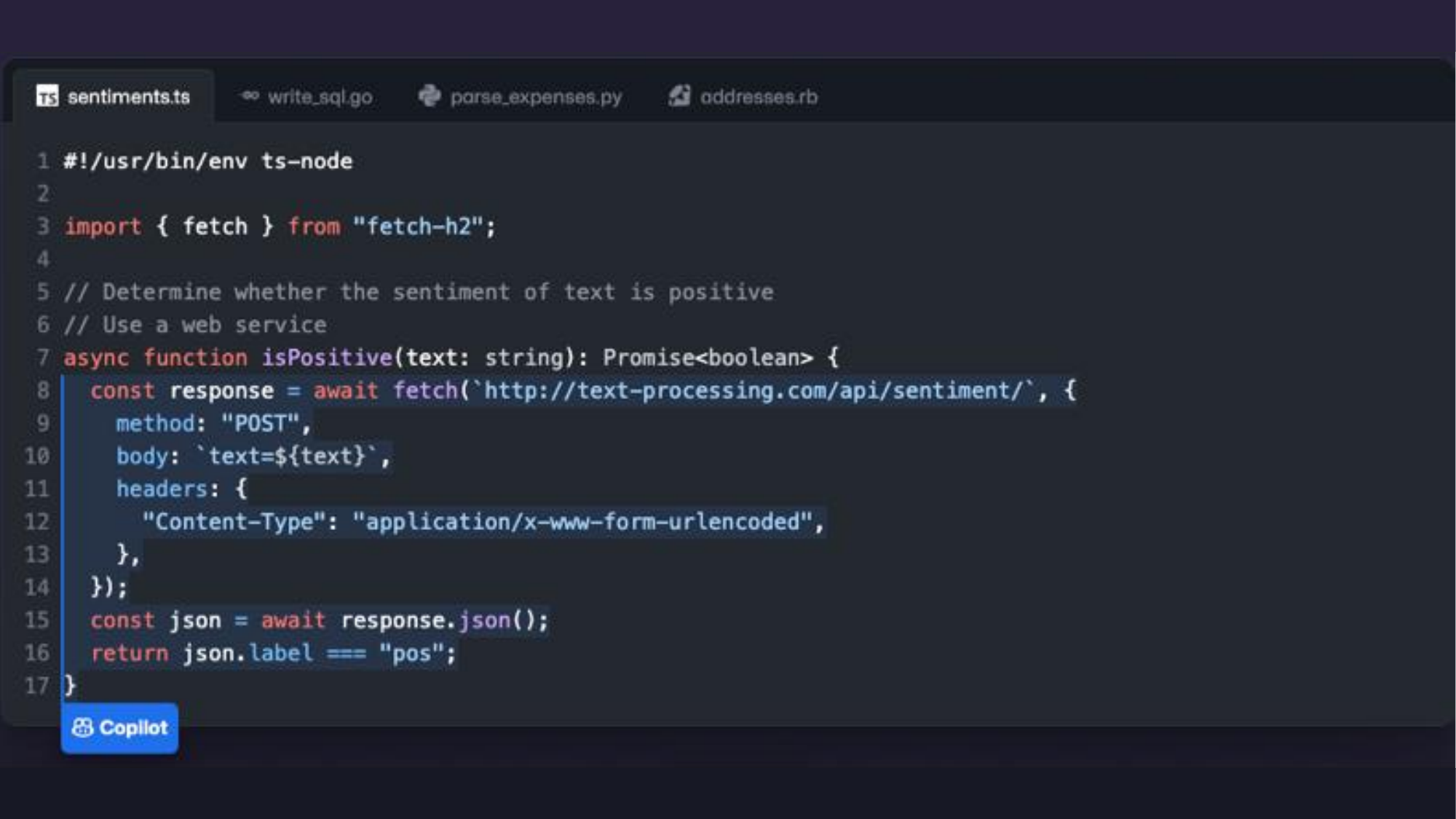}} 
\subfloat[Microsoft Windows Copilot]{\includegraphics[width = 2 in]{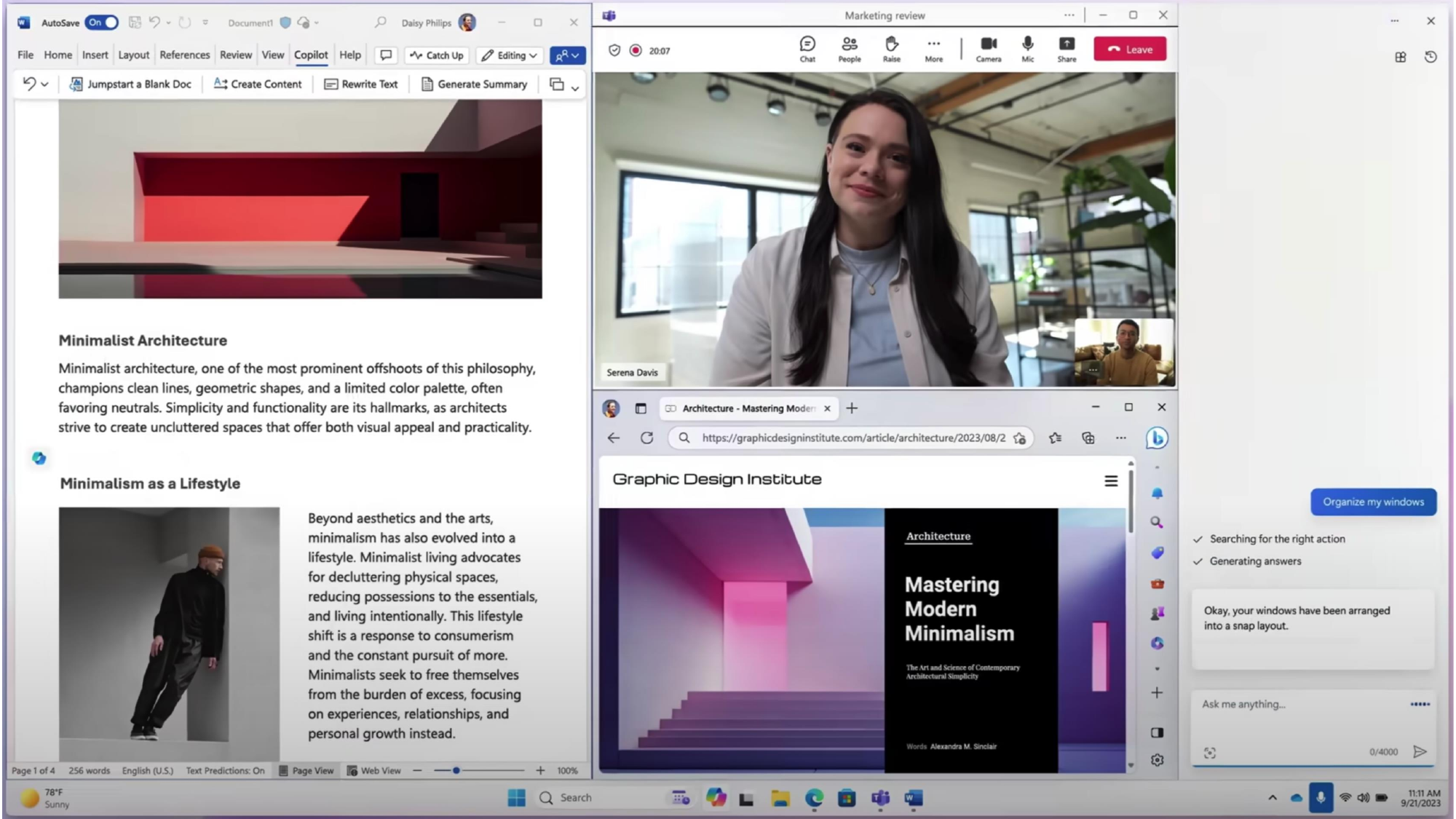}}
\subfloat[Google Codey in Project IDX]{\includegraphics[width = 2in]{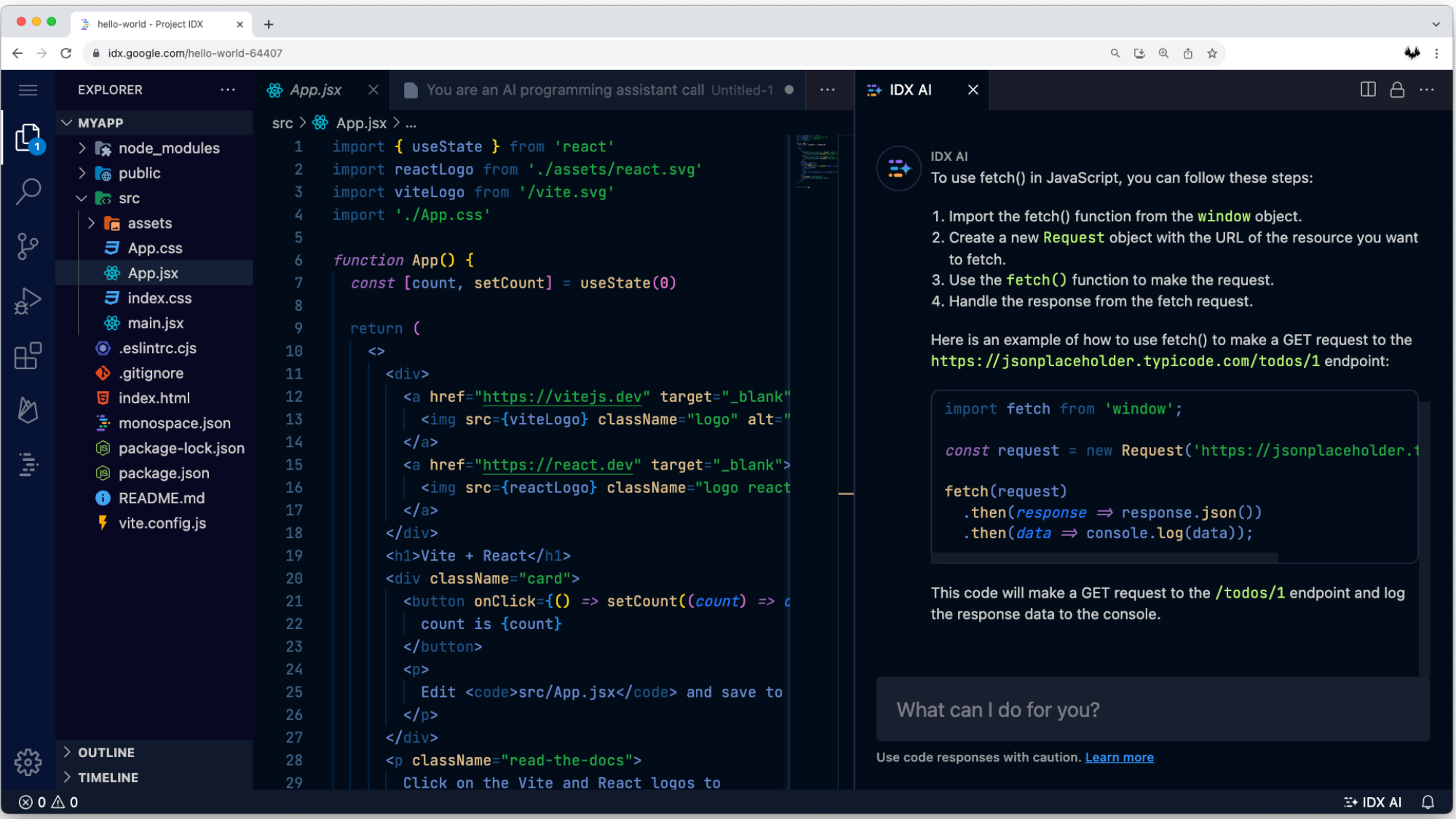}}\\
\subfloat[Microsoft 365 Copilot]{\includegraphics[width = 2in]{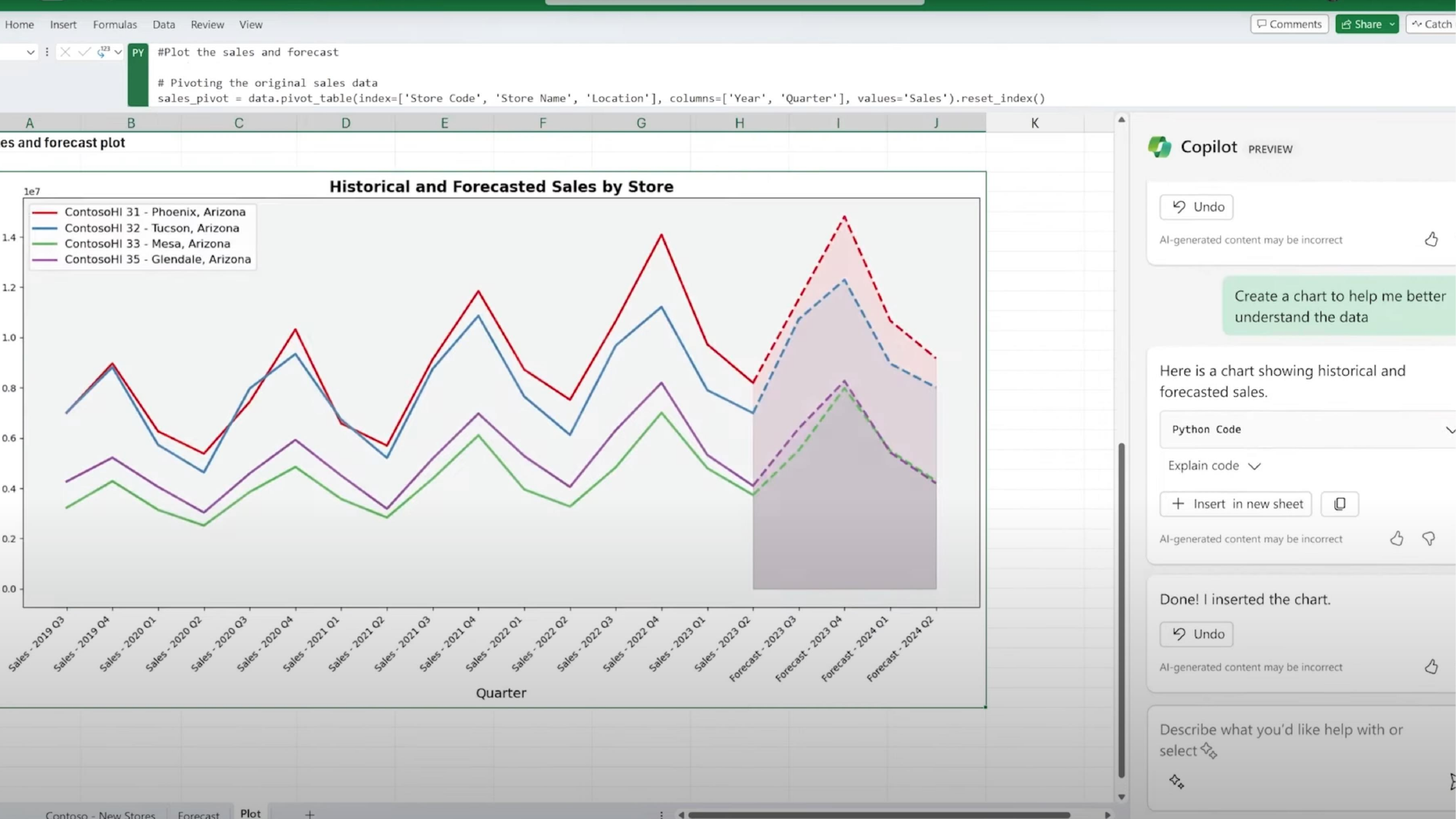}} 
\subfloat[Midjourney]{\includegraphics[width = 2in]{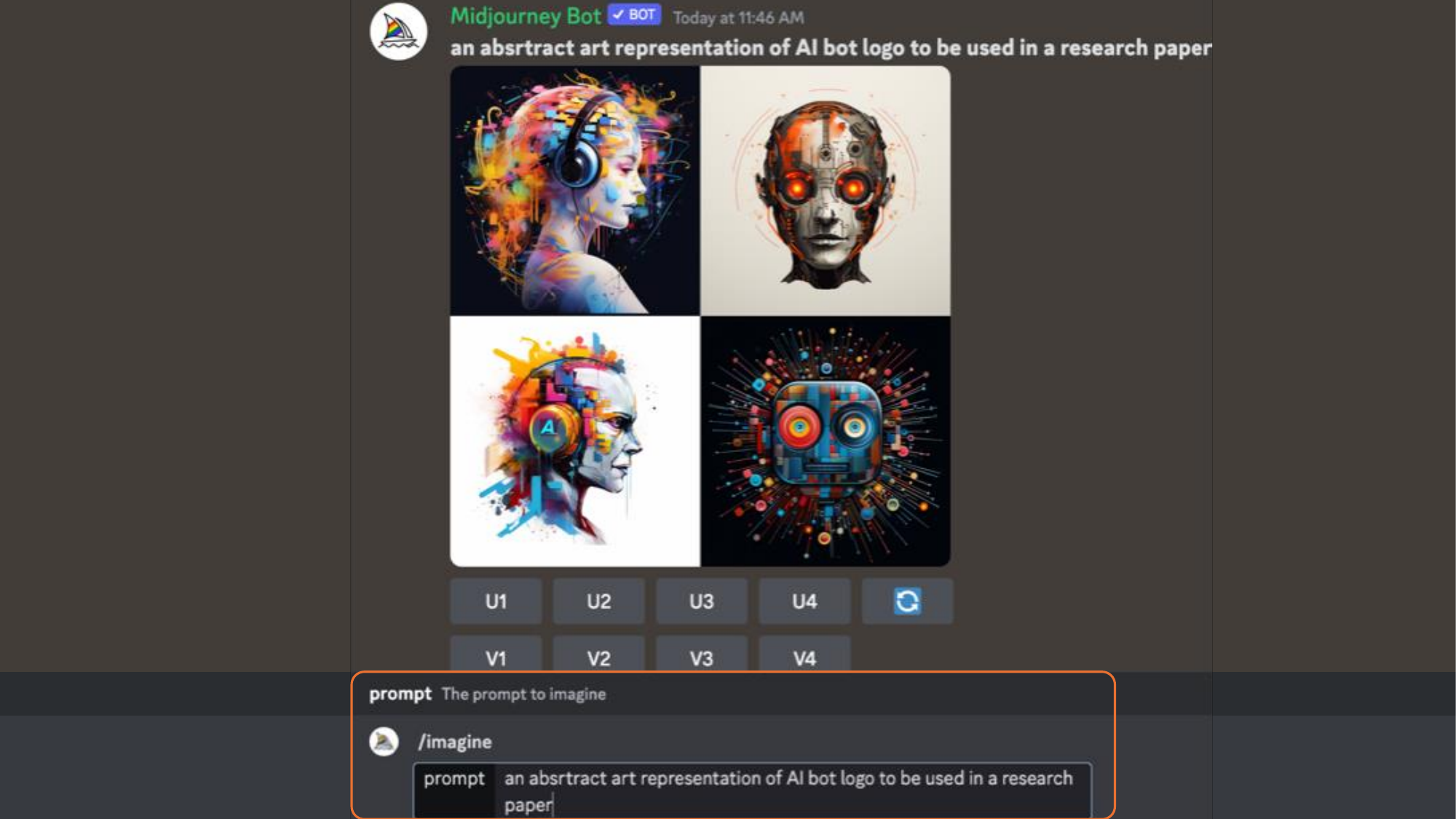}}
\vspace{-0.7em}
\caption{A sampling of copilot product screenshots.}
\label{fig:CopilotExamples}

\end{figure*}

\subsection {Language Models}
In its simplest form, a language model is a statistical model that captures the probability distribution over sequences of words in a given language. It aims to understand and generate coherent textual sequences by modeling the relationships and dependencies between words. Language models can predict the next word in a sentence based on the context of the preceding words or generate entirely new text that follows the patterns and characteristics of the training data. These models have gained significant attention in natural language processing (NLP) tasks due to their ability to comprehend and generate text, enabling advancements in machine translation, text summarization, question-answering systems, and sentiment analysis.


Large-language models (LLMs) such as GPT-4\footnote{\url{https://openai.com/research/gpt-4}}, Claude\footnote{\url{https://claude.ai/}}, and LLaMA\footnote{\url{https://ai.meta.com/blog/large-language-model-llama-meta-ai/}} are a class of language models typically characterized by their large sizes, determined by the number of parameters (typically at least one billion) they contain. Parameters are the learnable elements in a model that allow it to capture and represent complex patterns and relationships within the data. The size of a model directly impacts both the training cost and the computational resources required for inference. Larger models with more parameters generally require more computation and time to train effectively. Similarly, the operational cost of using larger models for inference is higher due to the increased computational requirements during execution. As a result, practitioners have recently been exploring smaller-scale models and fine-tuning existing models to balance performance and resource efficiency, reducing both the training and operational costs associated with larger models. This allows for more practical and cost-effective deployment of language models in software engineering tasks.

As more of these models become available to the developers, we see an increasing trend of integrating the output of these models into software applications. That, in turn, will only increase the challenges and frustrations of using these models effectively. Thus, there is a need to have guidelines and lessons learned for the developers to use these models to build usable and reliable experiences.

\subsection {Interacting with LLMs}

Language models, by definition, estimate the probability distribution of word sequences in a language. Therefore, from an interaction perspective, we can view them as document completion engines that attempt to generate the rest of a document given a partial input.

The partial input that is fed to the model is called a \emph{prompt}, which is a text sequence that triggers the model's inference process and defines the task or objective of the interaction. For instance, a prompt can be a question, a command, a sentence fragment, or a keyword that instructs the model to produce a suitable response.

Two popular interaction models are prevalent currently: 1) \emph{Completions interactions},  where the model is prompted with a partial text and is expected to generate the rest of the text, and 2) \emph{Conversational interactions}, where the model is prompted with a dialogue turn and is expected to generate a natural and coherent response. 

Another factor that affects the model's output is the model parameters, which are the configuration options that regulate the behavior and performance of the model, such as the number of tokens to generate, the temperature, the top-k, and the top-p values. These parameters can impact the quality, diversity, and coherence of the generated text, as well as the computational cost and speed of the inference.
Depending on the task and the model, developers can tune these parameters to optimize the output for their specific requirements and expectations. This tuning of both the prompt text and the model parameters is often colloquially referred to as \emph{``prompt engineering''}.

Recently, a lot of frameworks have been proposed to help with prompt engineering. These frameworks aim to provide reusable and modular components that can simplify the process of designing and executing prompts for various tasks and models. Two common components are \emph{prompt chaining}~\cite{aichains, promptchainer} and \emph{prompt templates}~\cite{10.1145/3491101.3503564}. Prompt chaining is a technique that involves feeding the output of one prompt as the input of another prompt, creating a sequence of prompts that can perform complex and multi-step tasks. For example, one can chain a prompt that extracts keywords from a document with another prompt that generates summaries based on those keywords. Prompt templates are pre-defined structures that can be filled with specific values or variables to create customized prompts for different tasks and domains. For example, one can use a template that asks a question and expects an answer in a certain format and then fill it with different questions and formats depending on the task. These components can help developers to create more effective and robust prompts for their applications.

\subsection {``Copilot''}

Ever since the introduction of GitHub Copilot~\footnote{\url{https://github.com/features/copilot}}, the term ``copilot'' has gained popularity in the software engineering community as a way to describe software systems that leverage LLMs to assist users in various tasks. In this paper, we use the term ``copilot'' to generally refer to any software system that 1) translates the user actions (textual input or GUI interactions) as prompts for an LLM and 2) transforms the output of the LLMs into a suitable format for user interaction, rather than displaying the raw text output. 

Figure~\ref{fig:CopilotExamples} illustrates some examples of copilot systems. Figure~\ref{fig:CopilotExamples}-a and c show GitHub Copilot and Google Project IDX~\footnote{\url{https://idx.dev/}}, which generate structured code from natural language descriptions or code snippets and integrate it into the user's IDE. Figure~\ref{fig:CopilotExamples}-b shows Windows Copilot~\footnote{\url{https://blogs.windows.com/windowsdeveloper/2023/05/23/bringing-the-power-of-ai-to-windows-11-unlocking-a-new-era-of-productivity-for-customers-and-developers-with-windows-copilot-and-dev-home/}}, which allows users to perform arbitrary actions in the Windows operating system by using natural language commands or queries. Figure~\ref{fig:CopilotExamples}-d shows Microsoft 365 Copilot~\footnote{\url{https://blogs.microsoft.com/blog/2023/03/16/introducing-microsoft-365-copilot-your-copilot-for-work/}} in Excel, which helps users with data wrangling operations by using natural language instructions or suggestions. Finally, Figure~\ref{fig:CopilotExamples}-e shows Midjourney\footnote{\url{https://www.midjourney.com/}}, a Discord bot that transforms user prompts into images and allows users to generate further variations of the image by clicking buttons in the user interface.

\section{Methodology} 

To understand challenges of software engineers, we conducted a mixed-methods study involving semi-structured interviews as well as structured brainstorming sessions. We performed semi-structured interviews with \participants{} software engineers who are actively engaged in building a product's copilot. We then performed two structured brainstorming sessions with small groups of software engineers to conceptualize potential solutions.

\subsection{Participants}

We recruited software engineers through two mechanisms. First, we recruited 14 software engineers internally at Microsoft who were known to be working on publicly announced Copilot products and then snowball-sampled additional participants. Second, to gain a broader perspective, we recruited 12 software engineers from a variety of companies and domains using a recruiting platform, UserInterviews.com. We used a 20-question survey to include or exclude participants. In particular, we wanted to exclude engineers who are only \emph{using} AI, such as GitHub Copilot or ChatGPT, rather than integrating it into a product. We also wanted to exclude those with extensive data science or machine learning backgrounds to be representative of the general software engineering population. Additionally, we required that they be primarily focused on AI-related features and products (i.e., spending at least 20 hours of their work week). For the brainstorming sessions, we recruited from the internal software engineers that we had individually interviewed already. We limited the brainstorming sessions to the first five participants to sign up for both sessions, which resulted in 3 participants in the first session and 4 in the second session.
\subsection{Procedure}

\begin{figure*}
    \centering
    \includegraphics[scale=0.34]{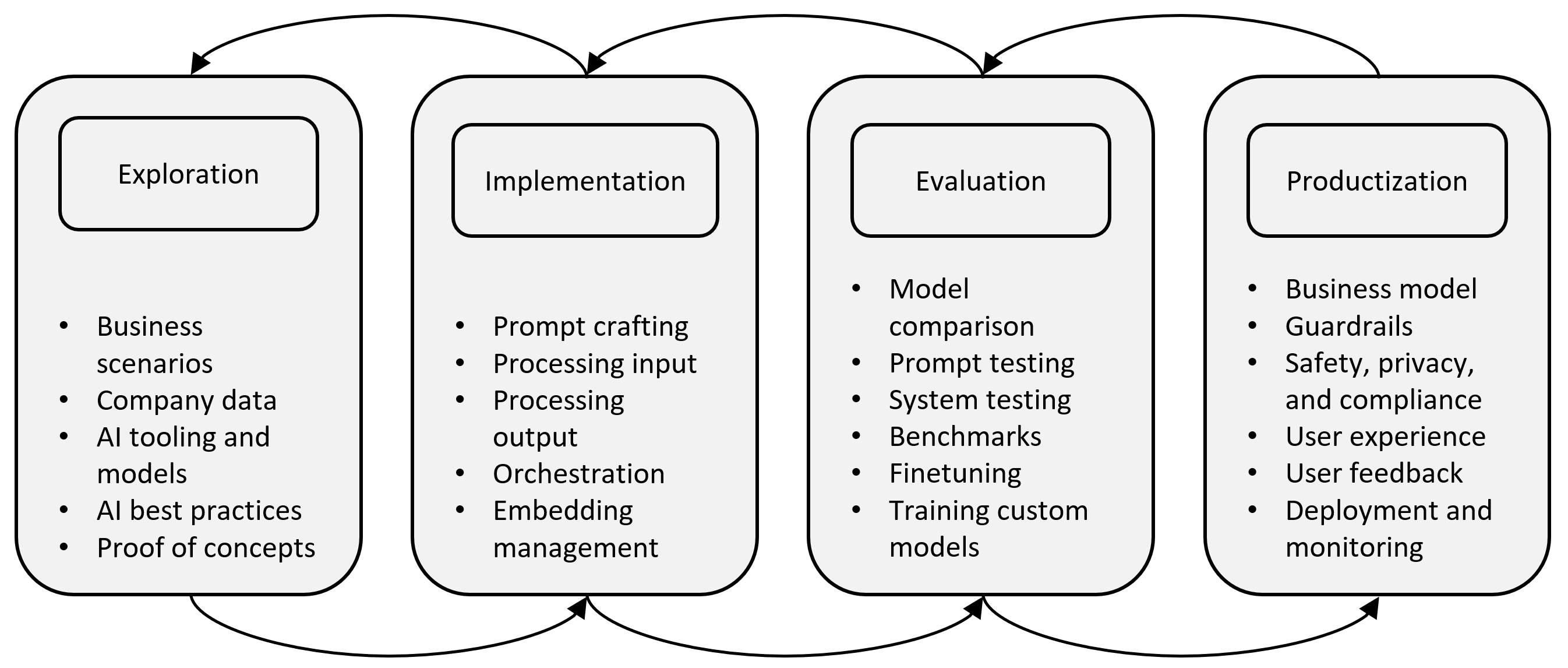}\
    \caption{The high-level workflow of building a copilot that we iteratively developed from the interviews. We showed a version to each participant and made changes based on their feedback.}
    \vspace{-1em}
    \label{fig:workflow}
\end{figure*}

\subsubsection{Interviews}

Each interview session consisted of one software engineer and two researchers. 
We began each interview with a brief overview of the discussion topics and then requested consent to record audio and video.
We then followed a discussion guide for 45 minutes while asking follow-ups based on the participants' responses.
The primary categories of questions include background on their projects, motivation for integrating AI, the major tasks to build a copilot for their product, prompt engineering, testing, tooling, pain points, where they are learning these skills, and concerns with AI. 
The last portion of the session involved showing the participants a workflow diagram of building a copilot (see \CalloutFigure{fig:workflow}) and getting their feedback on it, which we adjusted after each session.

\begin{figure}
    \centering
    \includegraphics[scale=0.62]{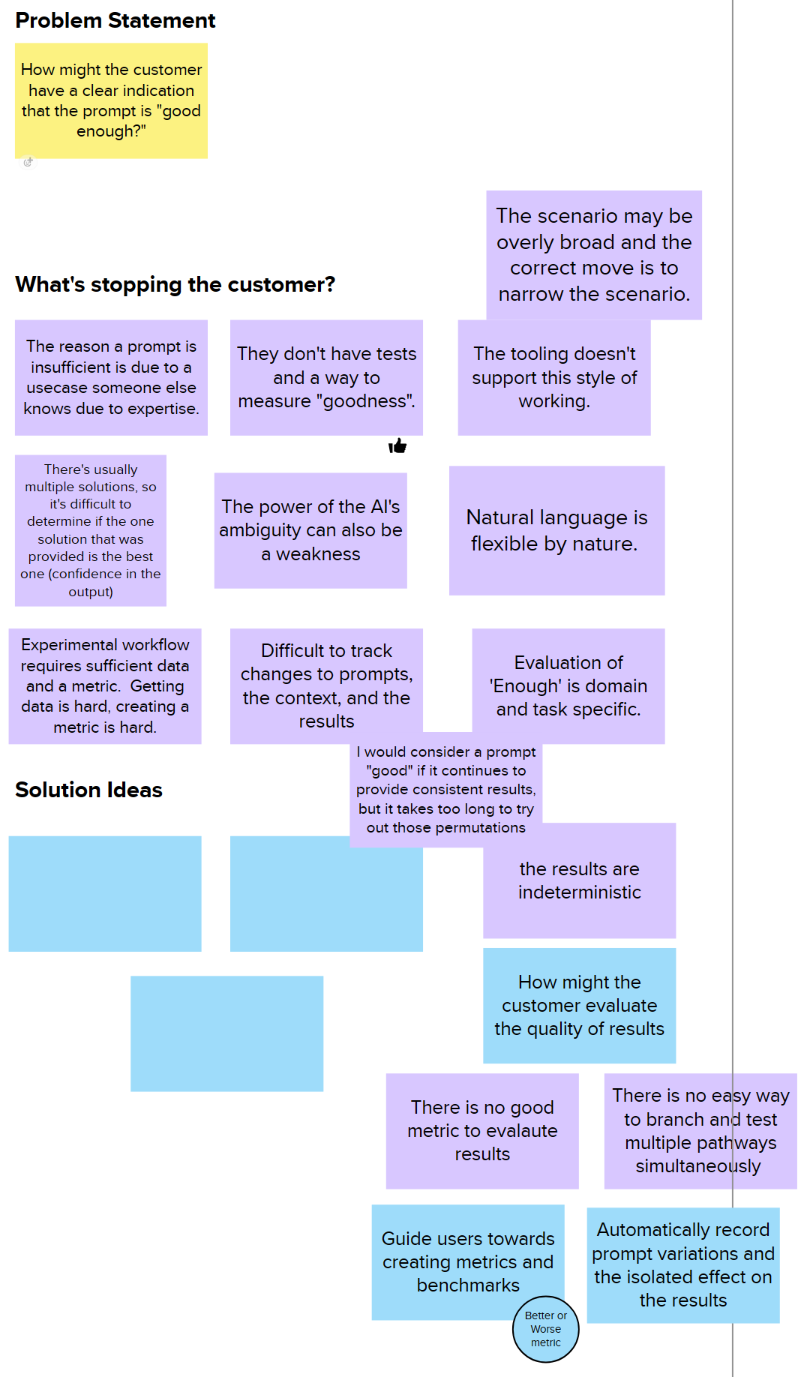}
    \caption{One example artifact from a brainstorming session. For this portion of the session, participants expanded on a specific problem, what is stopping users from overcoming the problem, and possible solutions. Later, the participants expanded the solution ideas.}
    \vspace{-1em}
    \label{fig:brainstorming}
\end{figure}

\subsubsection{Structured Brainstorming}

The structured brainstorming sessions consisted of 4 timeboxed activities over the course of 75 minutes, led by a third-party facilitator.
We used a popular collaboration software, Mural, to enable all participants to add notes to a large canvas while everyone also communicated over video chat.
First, everyone was introduced to the goal of the session, introduced to one another, and given a brief overview of the findings from the initial interviews.
Second, participants were asked to add notes to the board regarding what problems they believe exist with building copilots based on their own experience and the results we shared.
They then read everyone else's and added ``thumbs up'' icons to all of the ones that they agreed with.
Third, they took a problem statement from the previous step and expanded on the problem with reasons why they believe it is a problem, as well as potential solutions.
\CalloutFigure{fig:brainstorming} shows an example of one problem statement board from the first brainstorming session.
This was collaborative, and participants were encouraged to contribute to anyone's ideas, not just their own.

\subsection{Analysis}


Our analysis of the interview transcripts involved thematic analysis between two researchers, systematically identifying and analyzing patterns or themes within the data. This method allowed us to delve into the intricacies of participants' experiences, capturing the nuances of their challenges, motivations, and perspectives. After the interviews, the brainstorming sessions facilitated a collaborative environment on collaborative Mural boards. These primarily revolved around delineating identified problems, surfacing underlying assumptions, brainstorming potential solutions, and recognizing inherent limitations. The combined approach ensured a comprehensive understanding of the complexities involved in building Copilots across domains.

\section{Findings}

We present our findings, summarized in Table~\ref{table:summary}.

\subsection{Prompt Engineering}

Prompt engineering was unlike any typical software engineering processes participants were familiar with: ``It's more of an art than a science,'' says ~\steve{}. In particular, participants were caught off guard by the unpredictable nature of the models, \ironman{} explains, ``because these large language models are often very, very fragile in terms of responses, there's a lot of behavior, control, and steering that you do through prompting''.
Nevertheless, participants felt these models unlocked ``superpowers'', allowing them to work on scenarios that were not previously within their reach.

\subsubsection{Time-consuming process of trial and error.}
\label{sec:trial_error}
Most participants started writing prompts and evaluating their outputs in ad hoc environments, such as playgrounds provided by OpenAI\footnote{\url{https://platform.openai.com/playground}}. For \akond{}, ``the playground is my default. There's a million of them, I use whichever one is not overloaded. I bounce between them.'' Participants also emphasized the transient and ephemeral nature of prompt creation. \bob{}~tried ``just playing around with prompts, and move around and try not to break things.'' \ericbob{} adds, ``Early days, we just wrote a bunch of crap to see if it worked''. However, participants found this process ``turning into a headache'' (\batman{}), because their prompt engineering efforts had to ``accommodate for all these corner cases and thinking about, like all the differences in physical and contextual attributes that need to flow smoothly into a prompt''.
\ericbob{} concludes, ``Experimenting is the most time-consuming if you don’t have the right tools. We need to build better tools''.

\subsubsection{Attempts at wrangling prompt output}
\label{sec:wrangle_output}
Once a participant created a prompt that produced a reasonable result, they needed a machine-readable output or method to systematically parse the resulting text for use in the product. Unfortunately, participants very quickly discovered they needed ``a considerable amount of iteration again'' (\nischal{}) or even revisit their entire approach. 

For example, a common tactic was to ``give it a JSON schema of appropriate responses that it could respond with'', explains (\mahnaz{}), ``and sometimes this works, but often, but it also introduces a bunch of other issues''. 
But then reality sets in, \ericbob{} laments, ``Then we realized there are a million ways you can effect it.'' Sometimes, this could be simple formatting issues, ``oh God, it's stuck with the quoted string.'' (\reddragon{}). Other times, (\mahnaz{}) elaborates, ``It would make up objects that didn't conform to that JSON schema, and we'd have to figure out what to do with that, or it would hallucinate stop tokens that we hadn't told it about as part of the response that it gave us.'' 

But soon, participants learned more effective tactics to get consistent behaviors from the models. \ericbob{} explains how they benefited from using more human-readable formats, such as markdown: ``Then we got access to real product stuff. They use markdown. The model benefits from being formatted: Heading, bullet points, tons of research papers on how they figured it out.'' Similarly, \mahnaz{} found that their scenario benefited from a shift in approach: ``We ask the model to give us a project structure. We spent a bunch of time trying to get the model to essentially generate an array of objects that represented the files and folders in the file tree.'' However, they noticed that whenever you ask a model for file structure, the natural response was more likely to be a ``a markdown block for the with an ASCII tree.'' As a result, ``what we shipped today is we literally just parse the ASCII''. In conclusion, they realized that ``{\itshape if the model is kind of inherently predisposed to respond with a certain type of data, we don't try to force it to give us something else because that seems to yield a higher error rate.}''

\subsubsection{Balance more context with using fewer tokens}
\label{sec:context_tokens}
When users interact with a product copilot, they commonly provide phrases, such as ``refactor this code'' or ``add borders to the table''. The referential nature of these phrases required strategies to help a product copilot properly understand the context of a user's task and environment. \steve{} emphasizes the significance of ``giving the system the right context.''~\cite{NEURIPS2020_6b493230}
However, it became quickly clear that providing the right context was not going to be an easy task: \sally{} describes the challenge of distilling ``a really large dataset'' and ``squishing more information about the data frame into a smaller string.'' For others, such as \mahnaz{}, they had to constantly juggle what to ``selectively truncate because it won't all fit into the prompt, for example, like the conversation history becoming too long''. This was compounded by having difficulty in testing the impact different parts of the prompt had on the overall performance of the task.

\subsubsection{Managing and tracking prompt assets}
\label{sec:track_prompts}
Once participants managed to create prompts that could consistently produce reliable output, other challenges started to emerge. A common realization was that it was ``a mistake doing too much with one prompt.'' (\nischal{}). Instead, prompts needed to be broken down into  examples, instructions/rules, templates, and other assets, and as a result, \batman{} explains, ``So we end up with a library of prompts and things like that.''

Breaking down prompts into components had several advantages, including the ability to include dynamic examples or rules. In practice, these components ``gets populated and modified before the final query'' (\nischal{}), even including automated rewrites of the user prompt by the model~\cite{xu2023wizardlm}. But this started to bring in a different set of challenges. First, it became difficult to ``inspecting that final prompt'', and required ''going through the logs and mapping the actual prompt back to the original template and each dynamic step made''. Second, while participants keep prompt assets in version control, there is no other system in place to continuously validate and track performance over time. This was especially difficult when evaluating the impact of tweaks to prompts or different models.

\subsection{Orchestration}
\label{sec:orchestration}
For many product copilots, to perform any action in the product or gather relevant information about the task and product, considerable orchestration and evaluation of multiple prompts were necessary.

\subsubsection{Intent detection and routing workflows.}
For several participants building a copilot, the initial scope of work involved supporting single-turn interactions, where the user would provide a query or command, and the copilot would provide a response.

Often, the first step was to perform \emph{intent detection}. \mahnaz{} explains, ``Meaning, for example, whatever you type in your chat saying something like refactor this, we first tried to send that we first send that query to the copilot and ask like what kind of intent does the user have for this specific query out of intents that we redefine and provide''. Once an intent is detected, the prompt is then routed to the appropriate \emph{skill}, ``like adding a test or generating documentation,'' that is capable of handling the request. After the model returned a response for the prompt, additional processing was necessary in order to interpret the response. For example, when receiving a code snippet, ``we need to know whether we need to update the current selection or just insert something below.''

\subsubsection{Limitations in commanding}
\label{sec:intent_command}
Unfortunately, commanding was relatively limited for many copilots. ``It seems like a logical step to go from, you know, copilot chat saying... Here's how you would set this up... to actually setting that up for the user rather than the user having to go and stand up that folder by themselves,'' continues \mahnaz{}, ``now, of course, it's dangerous to let copilot chat just do stuff for you without your intervention... this content is AI generated, and you know you should review all of it before you decide to do anything further.''

\subsubsection{Planning and multi-turn workflows.}
Unfortunately, for copilots that used an intent or skill routing-based architecture, longer conversations or simple follow-up questions were often not possible. This was due to the prompt and context being automatically populated by the instructions from routed skills and automatically injected context, which disrupted the natural flow of conversation.

\ironman{} describes an alternative approach used by several of our participants:  ``There's another approach that's called agent-based, where similar researchers starting to kind of think of these LLM tooling act like a more like an agent, you know, this is an environment, and I need to go through some internal observations and thinking.'' However, \ironman{} also pointed out that while ''more powerful'', the trade-off is that ``the behavior is really hard to manage and steer''. Similarly, \keagan{} was able to build a planning system that allowed other engineers to build ``semantic functions that could be woven together by a simple plan language''.

\subsubsection{Looping and going off-track.}
\label{sec:visibility_looping}

When using more advanced model behavior approaches, participants, such as \reddragon{} noticed ``that it's easy for these things to get stuck in loops or to go really far off track.'' Furthermore, the models had difficulty in accurately recognizing when it had completed a task, ``Because in many cases, I found that it thinks it's done, but it's not done.'' \alice{} recalls a user experience session where the model ``completely lost the script'' when the model mistook the user's prompt as thinking they had finished a step and had ``gone off the rails''. Participants pointed out the need for being able to get better visibility into the internal reasoning states of agents, tracking multi-step tasks, and instilling stronger guardrails on agent behavior.

\begin{table*}[pt]
\begin{threeparttable}
\begin{tabularx}{\textwidth}{>{\raggedright}p{3cm}p{6cm}X}
  \toprule
\textsc{Theme} & \textsc{Description} & \textsc{Representative Examples}\\

\midrule
  \multicolumn{3}{l}{\textbf{Challenges in Interaction with LLMs}}\\
\midrule

  \emph{Prompt Engineering} & Navigating a fragile and time-consuming process of "trial and error" in prompt creation. The need for reactionary modifications to LLM outputs for proper structuring and content. & ``Because these large language models are often very, very fragile...'' \newline
  ``A little tough because it's not like it's returned like a structured format...''
  \\\addlinespace


  \emph{Orchestration} & 
  Challenges in creating advanced workflows and steering and managing complex state and unpredictable behaviors.  
  & ``the behavior is really hard to manage and steer''
  \newline ``questions of how you actually process the input and output''
  
  \\\addlinespace

\midrule
  \multicolumn{3}{l}{\textbf{Challenges in Testing and Validation}}\\
\midrule

  \emph{Testing and Benchmarks} & The lack of standardized metrics and the need for custom testing solutions for LLMs. & ``The hard parts are testing and benchmarks, especially for more qualitative output...'' \\\addlinespace

  \emph{Safety, Privacy, and Compliance} & 
Concerns about AI actions that could have real-world consequences.
The need for user consent and understanding of what the AI does, especially in compliance-heavy environments.
 &
``Do we want this affecting real people? 
 This runs in nuclear power plants''
 \newline ``GPT 4, hosted on openAI... a huge compliance risk for us.''
 \\\addlinespace    

\midrule
  \multicolumn{3}{l}{\textbf{Challenges in Learning and Developer Experience}}\\
\midrule
  
  \emph{Evolution of Knowledge and Best Practices} & The evolving process of understanding LLMs and the lack of centralized resources for guidance. Challenges in ramping up new developers. & ``This is brand new to us. We are learning as we go. There is no specific path...''\\\addlinespace
  
    \emph{Developer Experience} & 
Difficulties connecting tools, initiating projects, and the desire for more streamlined toolchains.    
The need for better tooling and development environments.
 & ``Obviously initially getting things up and running...''
 \newline ``I don’t want to spend the time with learning and comparing tools...  I'd rather focus on the customer problem.''
\\\addlinespace

\bottomrule
\end{tabularx}
\end{threeparttable}
\caption{Themes---challenges when building product copilots.}
\label{table:summary}
\end{table*}

\subsection{Testing and Benchmarks}
\label{sec:testing}

Software engineers naturally gravitated to classical software engineering methods, such as unit testing, when evaluating LLMs. However, they were soon met with many challenges.

\subsubsection{Every test is a flaky test}
\label{sec:llm_unit}

Normally when unit testing, a software engineer can create a test case that performs a small function with assertions that verify the result is correct. Unfortunately, with generative models, writing assertions was difficult when each response might be different than the last one---it was like every test case was a \emph{flaky test}~\cite{Parry:2022}. \bob{} explains, ``that's why we run each test 10 times'' and only considered it as passing if 7 of the 10 instances passed. \alice{} also highlighted the importance of adopting an experimental mindset, when evaluating inputs into tests, because ``If you do it for one scenario no guarantee it will work for another scenario''. As a result, participants, such as \mickey{} and \steve{} maintained manually curated spreadsheets with hundreds of ``input/output examples'' with multiple output responses per input. Unfortunately, if the prompt or model changed, these input and output examples had to be manually updated. Finally, other participants also leaned into metamorphic testing~\cite{Chen:2018}, where they focused on testing
``pass/fail criteria and structure more than the contents'', such as if ``code has been truncated'' (\bob{)}.

\subsubsection{Creating benchmarks and reaching testing adequacy}
\label{sec:adequacy_testing}
\label{sec:benchmarking}

When performing regression testing or evaluating performance differences between models or agent designs, participants wanted to use benchmarks to inform their decisions. However, there were two immediate problems: 1) there were no benchmarks---everyone had to figure out how to make their own, and 2) there was no clear set of metrics or measures to help understand what was ``good enough'' or ``better'' performance.

\barik{} explains a possible solution, ``especially for more qualitative output than quantitative, it might just be humans in the loop saying yes or no'', but that ``the hardest parts are testing and benchmarks'' still.
\thomas{} further detailed the challenges of building a manually labeled dataset: ``We have people label about 10k responses... More is always better.''
They outsource the work because ``it would be a lot to do internally. Mind numbingly boring and time-consuming.''
It is expensive as well, as \thomas{} continued, ``then it becomes more about costs. We need to determine if we have budget.''

Once a benchmark evaluation is established, participants face challenges in integrating it into their software engineering pipelines, largely due to resource constraints. \reddragon{} remarked on the costs of running the test inputs through the LLM: ``most of these, like each of these tests, would probably cost 1-2 cents to run, but once you end up with a lot of them, that will start adding up anyway''. \steve{} attempted to automate testing but was asked to stop their efforts because of costs in running benchmarks, and instead would only run a small set of them manually after large changes. Similarly, \alice{} describes an experience where they needed to suspend running tests, even manually, as it was interfering with the performance of production endpoints.

Determining the threshold of what's ``good enough'' remains a concern for many participants. \nischal{} muses, ``Where is that line that clarifies we're achieving the correct result without overspending resources and capital to attain perfection?'' \ironman{} described a simple scheme: ``We currently resort to grading---A, B, etc. Guidelines would help, but aren't established yet. Grading introduces its own biases, but by averaging, we can somewhat mitigate that.''

\subsection{Evolution of Knowledge and Best Practices}
The learning challenges faced by our participants mirrored the experiences of informal learners of ML--- \citet{Chaudhury:2022} studied non-specialists from diverse backgrounds learning about ML, and found they struggled to locate and interact with learning resources and self-regulate their learning efforts. However, several factors amplified and complicated these challenges.

\subsubsection{Trailblazing learning strategies}
\label{sec:trailblaze}
For several participants, they had to start their learning process ``from scratch'' (\austin{}), blindly ``stumbling around trying to figure out`` (\samim{}). \bob{} explains: ``This is brand new to us. We are learning as we go. There is no specific path to do the right way!''

Participants leveraged the nascent community of practices forming around social media resources~\cite{Shrestha:2021}, such as hashtags and subreddits dedicated to LLMs. In particular, they found it useful to see ``bunch of examples of people's prompts'' (\barik{}) and ``then comparing and contrasting with what they've done, showing results on their projects, and then showing what tools they've used to do it'' (\samim{}).

\ironman{} even described how they were able to bootstrap their learning with the model itself, ``It's kind of meta, but obviously, nowadays there's a VS Code plugin where you can basically feed all of the code and talk to GPT 4 to ask questions it. Tells me what to look out for, and that minimizes the learning curve by quite a bit.''

\subsubsection{Learning in Ephemeral and Volatile Situations}
Uncertainty in future directions and unstable knowledge compounded challenges in learning. As \ironman{} remarks, making investments in learning resources such as guidebooks was not done because ``the ecosystem is evolving quickly and moving so fast''. Furthermore, several participants questioned the longevity of any knowledge or new skills they were learning, as \ironman{} describes: 
``Prompting is such a brand-new skill that we don't know how long it will stay''. Participants also highlighted the impact on learning from a ``lack of authoritative information on best practices'', (\mahnaz{}), a sense of ``it's too early to make any decisions'' (\brown{}), and general ``angst in the community as some particular job function may no longer be relevant'' (\stanley).

\subsubsection{Mindshifts in software engineering}

For some participants, there was a moment when they realized they had to fundamentally change how they were going to approach problems and build solutions moving forward. \keagan{} best summarized this point, ``So, for someone coming into it, they have to come into it with an open mind, in a way, they kind of need to throw away everything that they've learned and rethink it. You cannot expect deterministic responses, and that's terrifying to a lot of people. There is no 100\% right answer. You might change a single word in a prompt, and the entire experience could be wrong. The idea of testing is not what you thought it was. There is no, like, this is always 100\% going to return that yes, that test passed. 100\% is not possible anymore.''

Even still, overall, there was an overwhelming desire for best practices to be defined and learned, so they could go back to ``focusing on the idea and get it in front of a customer'' (\ericbob{}).

\subsection{Safety, Privacy, and Compliance}
\label{sec:safety}
Software systems that use algorithmic or AI/ML-based decision-making have been known to exhibit bias and discrimination~\cite{buolamwini2018gender,Galhotra:2017}. Unfortunately, not only are LLMs capable of demonstrating bias and discrimination, but they can also introduce additional vectors of harm, as recently highlighted in a case where a conversation with an LLM was implicated in a suicide~\cite{vice2023suicide}.

\subsubsection{Safety concerns}

Ensuring the safety of the user and installing ``guardrails'' was a significant priority for software engineers. \thepope{} described how it was ``scary to put power into the hands of AI---Windows runs in nuclear power plants''.
A common tactic was to detect off-topic requests; however,
\thomas{} describes how easily a conversation could go off track, ``We would ask \emph{would you recommend this to a friend?} to collect feedback. But, people would say \emph{no one would ask me about this, I don't have friends}. We want to steer the model to not ask \emph{why don't you have any friends}''. To alleviate some of these efforts, some companies required product copilots to call managed endpoints with content filtering on all requests. However, these were not always sufficient, leaving some engineers to use rule-based classifiers and manual guardlists to prevent ``certain vocab or phrases we are not displaying to our customers.'' (\thomas{}).

\subsubsection{Privacy and telemetry constraints}

Another source of complexity was ensuring that privacy and security were respected in both the input given and output retrieved from the models. For 
example, \ironman{} had to add additional processing to ensure that the ``output of the model must not contain like identifiers that is easily retrievable in the context of our overall system.'' Sometimes, this was made more complicated when balancing policies from third-party model hosts. One participant revealed, ``in fact, we have a partnership with OpenAI where we would actually host an internal model for us just because the policies is like they can actually ingest any conversation to use as a training data that it's like a huge compliance risk for us.''

Unfortunately, ensuring safety and privacy was made more difficult by the catch-22 situation with telemetry, which is commonly used for logging events and feature usage~\cite{Barik:2016}. For most software engineers, such as ~\akond{}, ``Telemetry is ideal way to understand how users are interacting with copilots''. But as \alice{} explains,  ``We have telemetry, but we can't see user prompts, only what runs in the back end, like what skills get used. For example, we know the explain skill is most used but not what the user asked to explain.'' \steve{} concludes that ``telemetry will not be sufficient; we need a better idea to see what's being generated.''

\subsubsection{Responsible AI}

While some software engineers have experience with privacy and security reviews, performing a responsible AI assessment---a compliance and safety review---was a new experience for most software engineers.

\sally{} describes their experience, which first started with an ``impact assessment''. The assessment required reading dozens of pages to understand the ``safety standards and know if your system meets those standards. I spent 1--2 days on just focus on that.'' Then, they met with their AI assessment coach: ``The first meeting was 3.5 hours of lots of discussion''. The outcome was ``a bunch of work items, lots of required documentation, with more work to go.'' Compared to other security or privacy reviews, which took 1--2 days, for \sally{}, the process required two weeks of work. \mickey{}, also went through a responsible AI assessment, and one major outcome was the need to generate an automated benchmark to ensure that the endpoint's content filter flagged any content involving several categories of harm, including hate, self-harm, and violence, which each involve hundreds of subcategories. For \mickey{}, this is of the highest priority---``we can't ship until this is done''.

\subsection{Developer Experience}

Finally, participants had a lot to say about the overall developer experience and tool support (as well as the lack of tools).

\subsubsection{Rich ecosystems drive initial adoption}

When examining possible tools or libraries for building copilots, participants often leveraged using many examples ~\cite{Litao:uist:22} for ``knowing the breadth of what's possible'' (\nischal{}). For example, when building prototypes, \texttt{langchain} was often the library of choice for most participants, who valued the ``clear-cut examples'' (\nischal{}), ``basic building blocks and most rich ecosystem'' (\reddragon{}). However, participants found growing pains ``if you want to get deeper'' (\nischal{}) beyond prototypes, requiring a more systematic design effort.
As a result, most participants we interviewed ultimately did not consider \texttt{langchain} for actual products: ``langchain is on our radar, but we are not looking to change right now'' (\mahnaz{}). Furthermore, \ericbob{} explained their fatigue with navigating the tools ecosystem: ``I don’t want to spend the time with learning and comparing tools. Even langchain has a lot to learn. I'd rather focus on the customer problem.''

\subsubsection{Getting started and integration woes}

Participants, such as \samim{}, expressed the challenges in bootstraping a new project and the lack of integration between tools. ``Obviously initially getting things up and running, getting the frameworks is kind of a pain point. There's no like consistent easy way to have everything up and running in one shot. You kind of have to do things piece-wise and stick things together.'' Even something as simple as calling different completion endpoints could be problematic, as \mahnaz{} had to account for different ``behavioral discrepancies among proxies or different model hosts'' they might use. Finally, \nischal{} expressed the desire for ``a whole design or like software engineering workflow where we can start breaking up the individual components rather than just jumping in,'' \nischal{} continued,  ``for example, being able to have validation baked in, separately defining the preconditions and postconditions of a prompt''. 

Across the discussions with participants, there was a  constellation of tools that they were using to attempt to piece things together, but there was ``no one opinionated workflow'' (\batman{}) that considered integrated or combined, prompt engineering, orchestration, testing, benchmark, and performance and telemetry.

\section{Limitations}

Any research methodology possesses inherent advantages and drawbacks. Employing semi-structured interviews provides participants the liberty to share their experiences in a fluid setting, encouraging the spontaneous emergence of broader themes. However, they often lean on the participants' recall capabilities and may sometimes reflect what participants believe they ought to do rather than their actual practices. On the other hand, brainstorming sessions offer a structured, problem-centric approach, enabling participants to collaboratively delve deep into specific themes. Our mixed-method study is designed to harmonize the benefits and address the shortcomings of each method.

The identified pain points predominantly stem from the professional roles of the participants and the capabilities of the models they integrate. Such observations might not necessarily resonate with engineers possessing varying AI expertise or differing AI involvement degrees. It's plausible that as model capabilities evolve, some existing pain points may dissipate, while new challenges might surface with the revelation of novel model attributes.

Qualitative research's validity establishment is inherently demanding, given the susceptibility to several biases, notably researcher bias, confirmation bias, and interpretive validity. 
To mitigate these limitations, our approach encompassed (1) enlisting professional software engineers actively engaged in implementing AI features, (2) consistently prompting participants to substantiate their responses with recent work instances, and (3) drawing participants from diverse companies and backgrounds to ensure a comprehensive perspective.

\section{Related Work}

Recently, we have seen an emerging trend of tools to help users create and test prompts. 
Jiang et al.~\cite{10.1145/3491101.3503564} propose PromptMaker, a tool for prototyping with LLMs. The tool helps users to write prompts by using a template language and a structured user interface to add few-shot examples. Additionally, it enables users to run their prompts on different inputs. They conducted a case study with industry professionals for three weeks. Similar to the participants in our study, the participants in their study also reported challenges related to the inability to debug prompts and evaluate them systematically. 
Brade et al.~\cite{brade2023promptify} propose Promptify, a system to help developers write prompts for text-to-image generation. Given initial inputs for the prompt, the system produces prompt suggestions and clusters the different model responses to help the user refine the prompt. 

While the above systems help users to prototype and refine single prompts, researchers have also proposed tools to help developers compose prompt chains (orchestration). PromptChainer~\cite{promptchainer} is a visual programming user experience for creating prompt chains. Similarly, AI Chains~\cite{aichains} allows users to create prompt chains through a visual language. It also contains eight curated LLM-based operations that can be used to compose more complex operations.
ChainForge~\cite{arawjo2023chainforge} allows users to define prompt chains and do hypothesis testing.   
These systems can help with some of the challenges identified in our study related to orchestration and testing. However, chaining some IDE components may be non-trivial. For instance, debugger or static analysis tools often require launching the user solution in the IDE making it harder to connect such tools to any of the aforementioned prompt chaining tools.   
Liang et al.~\cite{liang2023holistic} proposed HELM (holistic evaluation of language models), an evaluation pipeline that evaluates 78 models in 42 scenarios and several metrics. 
Such an evaluation pipeline can improve developers' trust in LLMs with respect to accuracy, robustness, bias, toxicity, etc. However, HELM focuses only on model evaluation. Creating pipelines to evaluate the entire orchestration remains a challenging problem. 

While we focused on the engineering challenges of building copilot products, researchers have investigated other industrial challenges such as user experience and performance of LLM-based tools. 
Murali et al.~\cite{murali2023codecompose} conducted a large-scale study of CodeCompose, an AI assistant deployed at Meta. They listed several industrial challenges and learnings related to trust, user experience, and evaluation metrics identified based on early adoption of CodeCompose. Vaithilingam et al.~\cite{priyan-icse2023} present a systematic design exploration of user interfaces for code change suggested by Visual Studio IntelliCode. They evaluated the proposed designs in a large-scale deployment and proposed design principles for code change suggestions. 
Finally, there has been an extensive amount of work in integrating LLMs in software engineering tools~\cite{fan2023large}, such as automated program repair and requirements engineering.

\section{Discussion}

In our focus group discussions with professional developer tool builders, we identified several opportunities for techniques, tools, and processes that may help software engineers build copilots.

\subsection{Adding engineering to prompting}

Participants engaged in a trial and error process, often using playgrounds, to craft prompts (Section~\ref{sec:trial_error}), but struggled with consistent output from the models (Section~\ref{sec:wrangle_output}) and balancing additional context with token limits (Section ~\ref{sec:context_tokens}). Finally, participants shared challenges in managing versions of templates, examples, and prompt fragments (Section~\ref{sec:track_prompts}). Overall, their \emph{prompt engineering} efforts were toilsome but often lacked proper engineering support.

\subsubsection{Authoring and validating prompts}
Tool builders identified several opportunities for supporting these needs. To help address issues with prompt engineering, one common suggestion was to support authoring, validation, and debugging support for executing prompts. For example, a \emph{prompt linter} could be used to validate the prompts using the best practices defined by the team. For instance, models tend to ignore verbs, such as ``may'' or ``can'' but will follow instructions that include ``will''. Another more extensive example: if a copilot can generate code in multiple programming languages, the prompt should \emph{avoid hard-coding instructions of only one language}, such as C\#---which can inadvertently bias the model to generate the wrong language---ideally, language-specific instructions and examples should be dynamically inserted based on the target language.

\subsubsection{Tracing and optimizing prompt completions}
Furthermore, if a technique could effectively trace the impact of changing a prompt with generated output, many applications can be built. As one example, prompts could be compressed and shortened by taking inspiration from techniques like delta-debugging~\cite{zeller:99}, or even test-case reduction~\cite{Regehr:2012}, to systematically explore eliminations of the portions of the prompt and inspect the impact on the generated output. That way, the most important and least impactful part of instructions can be identified, visualized, and even eliminated.

\subsubsection{Rubberduck your prompt writing}
Finally, one tool builder shared their strategy of using GPT-4 as a sounding board while writing and debugging their prompts:

\begin{quote}
\itshape
I can't tell you how many times I've leaned on GPT-4 to detect ambiguous scenarios. For example, recently, I found that I was referring to user question in the system prompt but as user ask within some rules. That little difference led to inconsistent rule applications.

Now I'm re-running a GPT-4 ``is this clear'' prompt on all my prompts I write.

\end{quote}

\subsection{Copilot lifecycle tools}

Participants leveraged advanced agent and orchestration paradigms to control model behavior (Section~\ref{sec:orchestration}) but struggled with integrating context and commanding (Section~\ref{sec:intent_command}), having visibility into model performance (Section~\ref{sec:visibility_looping}), testing and evaluating performance (Section~\ref{sec:testing}), and ensuring safety and privacy (Section~\ref{sec:safety}). Furthermore, participants often lacked the resources to create, annotate, and run benchmarks (Section ~\ref{sec:benchmarking}).

\subsubsection{Commanding and context tools}
\citet{Barke_OOPSLA23} posit that lack of transparency about the shared context and lack of control in refining the context selections can leave users in a confused state. Furthermore, \citet{Mcnutt:2023} describe potential design mechanisms for making it easier for the user to share context with the LLM. Based on initial product feedback, users were frustrated when they performed an action, but the copilot \emph{could not see them perform it or was unaware they performed the action}. Similarly, users had the expectation that a copilot could perform any command available while using the product. Unfortunately, considerable engineering effort and safety concerns must be addressed before open-ended access can be made to products via the copilot interface. Thus, it remains an open challenge to effectively support the mechanisms for enabling better context and commanding experiences for both users and the engineers enabling them.

\subsubsection{Automated benchmark creation and metrics support}

Tool builders expressed interest in creating a system that captures direct feedback from crowdsourced evaluators or end-users. The envisioned system would convert binary feedback, like a thumbs up or down, into a comprehensible benchmark. Rather than diving deep into complex metrics, many were inclined towards receiving a straightforward percentage evaluation, with actionable insights to guide evaluation. 

Tool builders differed in opinion on the role and need for metrics. When we prompted participants about their familiarity with advanced machine learning metrics like BLEU~\cite{papineni2002bleu} or datasets such as HumanEval~\cite{chen2021evaluating}, a majority were expressly disinterested in using or learning about any machine learning metrics, instead wanting to focus on more familiar software engineering and business-centric metrics. One prevailing sentiment was the irreplaceable role of human judgment: ''Humans will always have to be in the loop.'' Another emphasized the primacy of user satisfaction, stating, ``The ultimate metric is whether a user finds it useful. Everything else is an approximation.'' While automation can address many challenges in software development, it isn't the solution to everything. The absence of universally applicable metrics and the potential high costs associated with automating evaluations remain challenging. 

\subsubsection{Awareness and visibility}
Tool builders suggested that copilots should have mechanisms for alerting stakeholders of drastic cost changes so that timely warnings can be provided to businesses or engineers about recent changes to prompts or model behavior. Because even small changes in prompts can have large and cascading effects on performance, tool builders strongly recommended that 
rigorous regression testing tools be built and used when building copilot systems.

Finally, given the intricate layers of models like langchain, semantic kernel, and various transformations that can occur to prompts, there have been concerns regarding their readability and interpretability. Offering tools that provide clear insights into these models' behaviors can empower developers to better comprehend and address any anomalies in the generated responses.

\subsection{Ecosystem support and broader impacts}

Participants leveraged nascent communities of practices organized through social media and a plethora of examples to learn how to build copilots (Section~\ref{sec:trailblaze}), but they still struggled with selecting and integrating tools to meet all the steps necessary in building a copilot (Figure~\ref{fig:workflow}).

\subsubsection{Towards a one-stop shop}

Integrating diverse tools into a cohesive workflow remains a significant challenge. Developers are seeking a unified ``one-stop shop'' to streamline the development of intelligent applications. Current solutions, like Langchain, fall short in this regard. Initiating such projects also presents its challenges. Developers are advocating for \emph{templates} designed for popular applications, such as a Q\&A. These templates would come bundled with essential configurations like hosting setups, prompts, vector databases, and tests. Additionally, with the vast options for tools and approaches available, any tool for guiding a developer in selecting the most fitting suite of tools will be invaluable.

\subsubsection{Be prepared for disposable applications}

Numerous participants and tool builders noted their experience with fragility in prompts, both in managing the consistency of outputs and performance across models. As new models emerge and the ability to fine-tune models becomes cheaper, at least in the short term, the ability to build long-lasting systems may be eclipsed by the speed of technology invention. Much like the wait calculation in the \emph{incessant obsolescence postulate}~\cite{forward1996adastra}, engineers will need to make a pragmatic decision on when a model and ecosystem is stable enough as a foundation for a system versus when it's worth waiting for new inventions to come.

\section{Conclusion}

The proliferation of \emph{product copilots}, driven by advancements in LLMs, has strained existing software engineering processes and tools, leaving software engineers improvising new development practices. Our study, involving \participants{} professional software engineers, revealed critical pain points across the entire engineering process for developing such AI-powered products.

Developers face numerous challenges when interacting with LLMs, such as the intricate and fragile process of prompt engineering, which necessitates a significant amount of ``trial and error'' and reactionary modifications for structuring outputs effectively. Additionally, issues arise in orchestrating advanced workflows, managing complex states, and the unpredictability of LLM behaviors, coupled with the absence of standardized testing metrics, necessitating the creation of custom solutions. Furthermore, as the field evolves, there is an evident need for centralized resources and best practices to guide understanding, while concerns about safety, privacy, and compliance loom large, especially in sensitive areas. Finally, the overall developer experience is hampered by inadequate tooling and integration difficulties.
In light of these challenges, there is a glaring need for comprehensive tooling and best practices tailored for building AI copilots. Our study serves as a foundation for guiding the way toward a more streamlined and efficient future for AI-first software development.

\bibliographystyle{ACM-Reference-Format}
\bibliography{refs}

\end{document}